\documentclass[11pt,a4paper,twocolumn]{article} 
\pdfoutput=1
\usepackage{my-jheppub}
\usepackage{amsmath, amssymb}

\usepackage{bm}
\usepackage{environ}
\usepackage{mathrsfs}
\usepackage{array,arydshln}
\usepackage{cuted}
\usepackage{comment}

\usepackage{graphicx,epsfig}
\usepackage{epic}
\usepackage{youngtab}
\usepackage{float}
\usepackage{color}
\definecolor{maroon}{rgb}{0.8,0.3,0.}

\usepackage{slashed}
\usepackage[nodayofweek]{date time}

\usepackage{hyperref}

\usepackage{aurical}
\usepackage[T1]{fontenc}

\allowdisplaybreaks

\newcommand{\be}{\begin{equation}}
\newcommand{\ee}{\end{equation}}

\newcommand{\bv}{\begin{verbatim}}
\newcommand{\ev}{\end{verbatim}}

\title{Reactive immunization on complex networks}
\author[a]{Eleonora Alfinito}
\author[b]{, Matteo Beccaria} 
\author[b]{, Alberto Fachechi} 
\author[b]{,  and  Guido Macorini} 

\abstract{\small{Epidemic spreading on complex networks depends on the topological structure as well as on the dynamical properties of the infection itself. Generally speaking, highly connected individuals play the role of hubs and are crucial to channel information across the network. On the other hand, static topological quantities measuring the connectivity structure are independent on the dynamical mechanisms of the infection. A natural question is therefore how to improve the topological analysis by some kind of dynamical information that may be extracted from the ongoing infection itself. In this spirit, we propose a novel vaccination scheme that exploits information from the details of the infection pattern at the moment when the vaccination strategy is applied. Numerical simulations of the infection process show that the proposed immunization strategy is  effective and robust on a wide class of complex networks.}
\vspace{0.2cm}\\ {\footnotesize PACS number(s): 89.75.-k, 87.23.Ge, 87.10.Rt}
\vfill }
  
\affiliation[a]{Dipartimento di Ingegneria dell'Innovazione,\\
Universit\`a del Salento, Campus Ecotekne, 73100 Lecce, 
Italy}

\affiliation[b]{Dipartimento di Matematica e Fisica Ennio De Giorgi,\\
Universit\`a del Salento \& INFN, Via Arnesano, 73100 Lecce, 
Italy}
                      
\emailAdd{eleonora.alfinito@unisalento.it} 
\emailAdd{matteo.beccaria@le.infn.it} 
\emailAdd{alberto.fachechi@le.infn.it} 
\emailAdd{macorini@nbi.ku.dk}

\notoc

\begin{document}


\twocolumn[
\begin{@twocolumnfalse}
\maketitle
\end{@twocolumnfalse}
]
\flushbottom

\section{Introduction}

Epidemic diffusion on complex networks \cite{newman2003structure,pastor2014epidemic,boccaletti2006complex} is a general paradigm to describe a large variety  of real world outbreaks of infections, ranging from the strictly 
biological case to malware diffusion as well as opinion propagation \cite{bornholdt2006handbook}. 
A central issue is the design of efficient immunization strategies able to
prevent or control the epidemic spreading  \cite{pastor2014epidemic}. 
In this context, numerical simulations are a flexible and well-controlled 
framework to study epidemic dynamics. In particular, they allow to understand the effectiveness of 
vaccination strategies that we shall broadly classify as preventive {\em vs.} reactive schemes.
{\it Preventive} immunization strategies aim to strengthen the network against 
epidemics using information about the {healthy} configuration, {\it i.e.} 
identifying the nodes to be immunized according to some  {\it score}
before the epidemic event. The score may require local or global knowledge about the network 
topological structure. 
An important example of the {preventive approach}
is the {\it Targeted Immunization} scheme (TI) \cite{pastor2002immunization} (see also \cite{chen2008finding, hebert2013global, yan2015global}),
originally designed for scale-free networks. 
The idea is to target nodes with high connectivity degree because they act as hubs in the infection spreading. 
A similar degree-based approach, but exploiting only local information, is the  {\it Acquaintance Immunization} (AI) \cite{cohen2003efficient}.
Some variations and improvements are discussed in \cite{stauffer2006dissemination,hu2012immunization}.

Instead, 
{\it reactive} immunization strategies start with the network undergoing a propagating infection
and take into account dynamical aspects of the network and of the epidemic itself to identify which are the best sites to be vaccinated.
Several scores have been designed considering, for instance, personal awareness about the epidemics \cite{ruan2012epidemic}, 
message-passing interactions \cite{altarelli2014containing}, dynamical reaction of the networks \cite{liu2014controlling,perra2012activity}, information from previous infections \cite{yan2014dynamical}, finite time for the vaccination to become effective \cite{pereira2015control},  {\it etc}.
A remarkably simple example of reactive protocols is the so-called {\it High-Risk Immunization} (HRI) \cite{nian2010efficient}, where 
the healthy neighbors of infected nodes are vaccinated.\par
In this paper, we propose a modification of TI scheme which exploits a refined score based on a local-global mixed strategy. Specifically, it introduces a modified score that is designed to consider both hubs and individuals at risk of contagion as relevant in the epidemic spreading. 
In other words, we attempt to use  the infection itself as a source of information and as a probe of
how the network reacts to the disease. On a regular network the infection may display a well defined
propagating front, then, a good strategy is to vaccinate in a neighborhood of it. It is not clear whether this strategy
makes sense on a complex network and we precisely try to answer this question.
The effectiveness of our strategy is tested by a Monte Carlo implementation of the SIR model \cite{kermack1927contribution,may1979population} on a variety of complex theoretical and real 
 networks {and systematically comparing our proposal with some standard immunization strategies \cite{pastor2002immunization,cohen2003efficient,nian2010efficient}}.

\section{Epidemics modeling and reactive immunizations: a new score}

The SIR model is a simple compartmental model of disease spreading \cite{kermack1927contribution}.  
Individuals are divided in three classes: susceptible ($S$), infected ($I$) and recovered ($R$).
The epidemic evolution is then modeled by the transitions  $S\rightarrow I$  and  $I\rightarrow R$.
In more details, it starts with  a single (patient zero) infected node.
Then, at each step of the Monte Carlo process, a randomly chosen infected individual can recover with probability $p_{\text{\tiny{SIR}}}$. 
Otherwise, one of its first neighbors is randomly selected and, if susceptible, gets infected. The reactive immunization takes place when a fraction $f$ (the {\it epidemic threshold}) of the population is infected.\footnote{{Due to the stochastic nature of the process,  epidemic may die out before reaching the threshold $f$
and immunization does not take place in these cases. The relation between the quantities $g$ and $\langle d_V\rangle$ is $\langle d_V\rangle = P_f g$, where $P_f$ is the probability that the infection reaches the threshold (which of course depends on the network and the threshold itself). We choose to take into account 
these events because they give an information about the exposure of a given network to a pandemic outbreak without vaccination. Given the value of $p_{\text{\tiny{SIR}}}$, the non-spreading events are relatively rare, for example for a BA[2] the probability to reach the lower threshold is roughly 90\%.}
} 
The vaccination is a single-step process in which a fraction $g$ of susceptibles individuals is immunized according to some score. 
The finite size of the network ensures that the system always reaches a steady final state without infected individuals. The density $d_R$ of recovered individuals in this state is clearly related to the spreading strength of the epidemic on the network. A good immunization strategy would therefore reduce the final density $d_R$ at the cost of a relatively low vaccinated density $d_V$. The average values $\langle d_R\rangle$ and $\langle d_V\rangle$ are computed by repeating the SIR evolution with vaccination a large number of times. \par
%
%
We propose a novel strategy of vaccination which interpolates between preventive and reactive immunizations. In doing so, we take into account both static information (like the network geometry) and dynamical information (like the pattern of a specific infection). To this aim, we consider the score
\be
\label{1.2}
\mathcal{S}_{i} = d_{i}+\,\sum_{j\in N_{i}}\bigg[
\beta \frac{\delta_{j, I}}{(d_{j})^{1/2}}
+\gamma\,\frac{\delta_{j, S}}{d_{i}}\,
\frac{d_{i}-d_{j}}{d_{i}+d_{j}}\bigg],
\ee
where $N_{i}$ denotes the set of 
neighbors of the $i$-th node, $d_i$ its degree ({\it i.e.} the number of links pointing to it), $\delta_{j,I}$ and $\delta_{j,S}$ are the Kronecker deltas which select only infected or suspectible neighbors
and $\beta,\gamma$ are  free parameters. 
We call our proposal  
{\it Locally-Modified Targeted Immunization} (LMTI$_{\beta, \gamma}$).
For $\beta=\gamma=0$, the score reduces to that of Targeted Immunization  \cite{pastor2002immunization}.
The $\beta$-term in the r.h.s. of (\ref{1.2}) favors the immunization of individuals {\it near} the epidemic front. 
The damping factor $(d_{j})^{-1/2}$ selects neighbors with lower connectivity, which constitute 
bottlenecks for the epidemic diffusion. It is therefore possible to reduce the contagion by cutting them off.
The $\gamma$-term is a further improvement involving the so-called {\it leverage centrality} \cite{joyce2010new}
restricted to the susceptible neighbors. It measures the reciprocal influence of the $i$-th node and its neighbors in the epidemic diffusion. {In fact, leverage centrality is a natural metric quantifying the local influence of a node on its neighbors and therefore it gives complementary local information with respect to the common (local) clustering coefficient.
}\par

We test the effectiveness of the score (\ref{1.2}) against the following benchmark immunization strategies
\begin{itemize}
\item{\textbf{Targeted Immunization (TI).}} Our implementation of TI follows the original idea: nodes are vaccinated according to their degree.
The only modification is that the immunization is performed as a reactive process when the epidemic reaches the threshold $f$.
Only  nodes yet susceptible at the vaccination time are protected.
\item{\textbf{Acquaintance Immunization (AI).}} As in the previous case, AC immunization \cite{cohen2003efficient} is implemented as a reactive process.
The choice of the nodes to be vaccinated follows the original proposal. 
Random first neighbors of randomly selected nodes are vaccinated (if susceptible) 	according to the desired immunized fraction $g$.
\item{\textbf{High Risk Immunization (HRI).}} Our implementation retains the  idea of \cite{nian2010efficient} to vaccinate neighbors of infected nodes, 
but the process is instantaneous and permanent. 
We test this strategy  by immunizing up to the $99\%$ of the first neighbors of the infected nodes at the vaccination time.
\end{itemize}
\section{Benchmark complex networks}

We test the effectiveness of our protocol on a variety of networks ranging from theoretical models to a selection of real networks.
In the first class, we consider the classical examples of Barab\`{a}si-Albert (BA) and Watts-Strogatz (WS) models.
The first one is the prototype of scale-free networks \cite{albert2002statistical,bornholdt2006handbook} and it is based on a growth algorithm 
with preferential attachment. We denote with BA[$Q$] the network built adding $Q$ new links at each step of the algorithm.
The second one is the prototype of small-world networks \cite{albert2002statistical,bornholdt2006handbook,j1998collective}. WS graphs are built starting from regular ones with $\mathcal{N}$ nodes (each one connected to $2Q$ consecutive sites) and then rewiring the links with probability $\theta$.
Here, we consider WS[$Q$] networks with $Q=2,3$ and $\theta=0.1$, $0.5$.

\begin{figure}[!htb]
  \centering
 \includegraphics[width=0.5\textwidth]{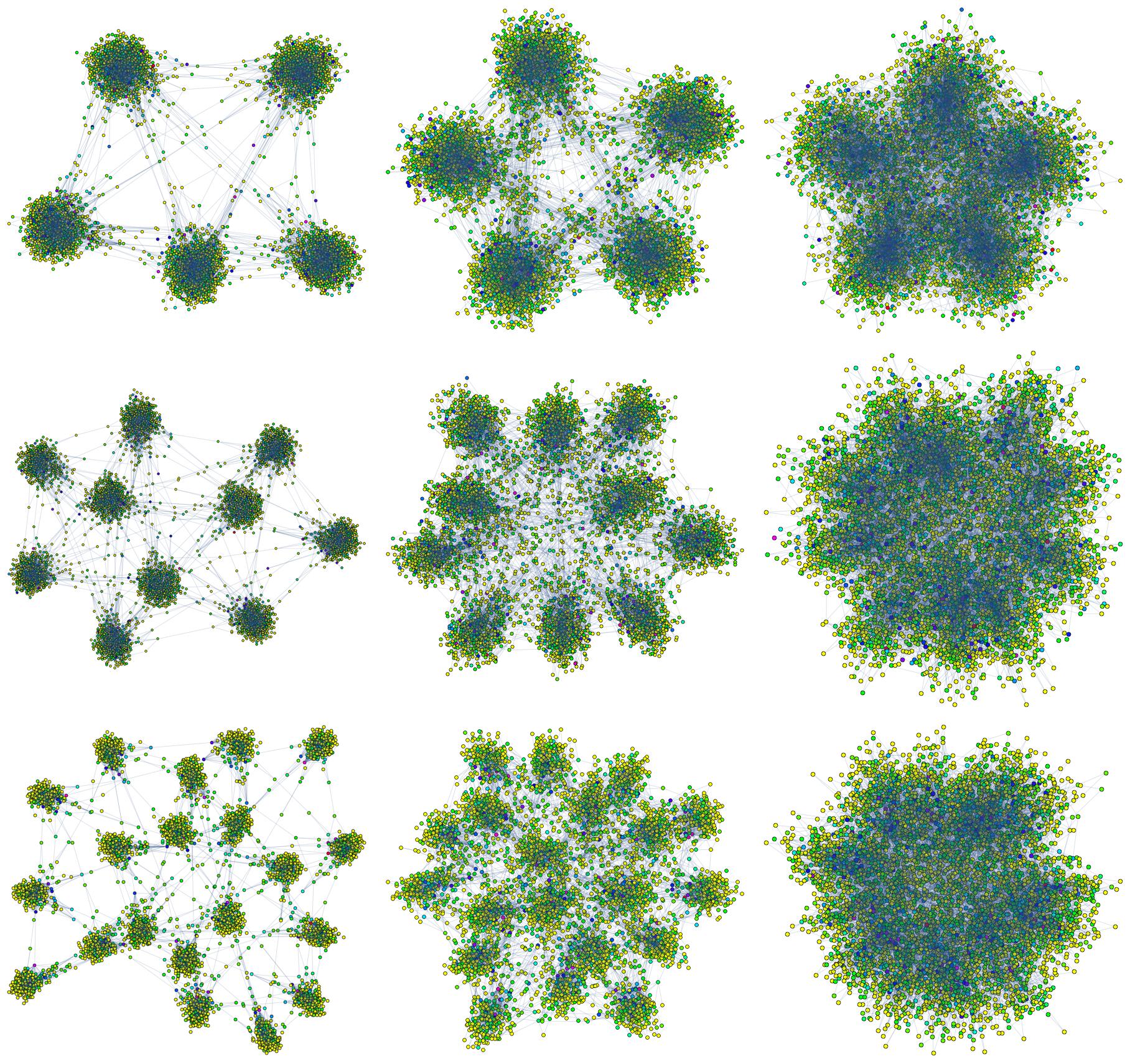}\vspace{-0.5cm}
   \caption[]
   {\label{fig:villi} Some examples of randomly connected BA networks. The first line shows $5$ BAs connected with 100 and 500 and 2000 random extra links. In the second and third line  
   the plots show examples of the networks obtained starting with 10 and 20 centers. 
   }
\end{figure}		

We also propose two modifications of BA model. 
The first one is based on a partial randomization procedure. We start with a standard BA[$Q$] network with $\mathcal N$ nodes, then we randomly rewire $\mathcal R$ links. 
In our tests, we consider $Q=2$, $\mathcal{N}=1000$ and $\mathcal R=100,500,1000,2000$.
The second variant is realized starting with $m$ disconnected BA[2] {\it centers}, further connected 
adding $k$ random links between nodes belonging to different BAs. 
Here, we consider a starting network with $\mathcal N=5000$ nodes, equally distributed in $m=5,10,20$ initial clusters, and $k=100,500,2000$.
This variant can be thought as a toy model for the epidemic spreading in clustered communities with relatively loose links. 
Some examples of the resulting networks are shown in Fig. \ref{fig:villi}.
\par

Besides these theoretical models, we also consider the epidemic spreading 
in the following real networks:

\begin{enumerate}
\item { {\sf {Internet\_AS}}, 11174 nodes, 23408 links.} It describes the undirected unweighted Internet Network\footnote{\url{https://sites.google.com/site/cxnets/research222}} \cite{colizza2006detecting}
at the Autonomous System level. The data were collected by the Oregon Route Views Project \url{http://www.routeviews.org/} in May 2001. 
Nodes represent Internet service providers and edges connections among them. 

 \item { {\sf {AA}}, 1057 nodes, 2502 links.}
It  describes the interactions between the metabolites of E. coli in the course of 
the metabolic cycle\footnote{\url{http://www3.nd.edu/~networks/resources/metabolic/}} \cite{jeong2000large}.
We consider the {\sf {AA}} case.
 
 \item { {\sf {CA-HepTh-pruned}}, 8638 nodes, 24836 links.}
The Arxiv HEP-TH (High Energy Physics - Theory) collaboration network\footnote{\url{http://snap.stanford.edu/data/ca-HepTh.html}}
from the e-print arXiv. 
A paper generates a completely connected subgraph in which nodes represent its authors.

\item { {\sf {p2p-Gnutella08}},  6300 nodes,  20776 links.} 
It is a sequence of snapshots of the Gnutella peer-to-peer file sharing network from August 2002.\footnote{\url{http://snap.stanford.edu/data/p2p-Gnutella08.html}}
Nodes represent hosts in the Gnutella network  and edges are connections among them.
  
\item {{\sf {ProteinYeast}},  1870 nodes,  2350 links.}
It is the Protein Interaction Network\footnote{\url{http://www3.nd.edu/~networks/resources/protein/bo.dat.gz} } \cite{jeong2001lethality}.

\end{enumerate}
To provide some additional informations, in Tab. \ref{tabella} we report the global clustering coefficients and mean distances among the nodes for the above real networks, and a comparison
with the same quantities computed for random networks.

For BA and WS models, we consider $50$ different realizations for each network and perform $10^5$ Monte Carlo runs with different initial conditions for each of them. 
For the BA variants, we consider $20$ different realizations of each graph and average $10^4$ runs for each one. 
Finally, for real networks the statistics varies from $10^4$ and $10^5$ runs, depending on their size. 
{With such a choice, we keep the statistical error on the final recovered density $\langle d_R \rangle$ under control (for instance, it is of the order of 0.1\% in theoretical models).}\footnote{{Statistical fluctuations are mainly determined by the simulation length, \textit{i.e.} by the number of MC steps, while the dependence on the particular network realization is rather weak due to self-averaging.}} In all cases, we fix the recovering probability to \mbox{$p_{\text{\tiny{SIR}}}=0.1$} and consider two epidemic thresholds  
 $f=0.05$ or \mbox{$f=0.15$}.\footnote{By comparison, in a  regular square lattice the epidemic threshold is $p_{c,\text{\tiny{SIR}}}=0.1765$  \cite{sirtome}, so 
$p_{\text{\tiny{SIR}}}=0.1$ would be in the spreading phase. {In this work, our main goal is a comparison of the relative effectiveness of the various vaccination strategies. A change in $p_{\text{\tiny{SIR}}}$ will surely affect the final balance of the epidemic, but, from the point of view of the comparison of the strategies, the dependence on $p_{\text{\tiny{SIR}}}$ is not crucial. Provided that $p_{\text{\tiny{SIR}}}$ is low enough to give a spreading epidemic, a change of the value of the recovering probability results in an overall shift of all the curves, but does not change the relative performances.}}

\section{Results}
In this section, we report the main results of our Monte Carlo simulations. In particular, we compare the various immunization strategies according to their ability in reducing the epidemic prevalence $\langle d_R\rangle$ by $50\%$ and $75\%$ (the horizontal dotted lines in the plots) and in reaching the epidemic threshold (red solid line in the plots).\par
\begin{figure}[!hbt]
    \centering
      \includegraphics[width=1.0\columnwidth]{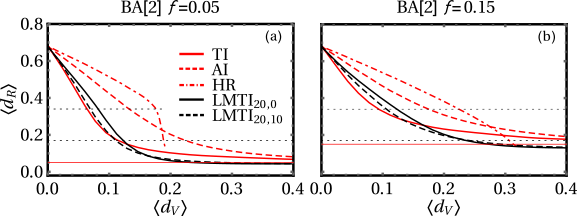}\vspace{-0.5cm}
      \caption{The recovered mean final density $\langle d_R \rangle$ as a function of the mean fraction of vaccinated $\langle d_V \rangle$, for BA[2] with $\mathcal N=1000$ nodes.
The LMTI scheme is compared to TI, AI and HR immunization strategies. The horizontal red solid line is the epidemic threshold, $f=0.05$ (a), $0.15$ (b), while the horizontal dotted lines are $25\%$ and $50\%$ of the mean final density of recovered without any vaccination.}\label{ba}
\end{figure}
\begin{figure}
    \centering
     \includegraphics[width=1.0\columnwidth]{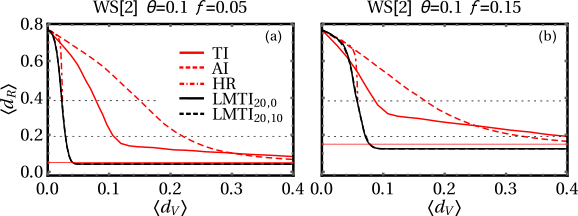}\vspace{-0.5cm}
    \caption{The recovered mean final density $\langle d_R \rangle$ as a function of the mean fraction of vaccinated $\langle d_V \rangle$, for WS[2] with $\mathcal N=1000$ nodes and the rewiring probability $\theta=0.1$.
		The LMTI scheme is compared to TI, AI and HR immunization strategies. The horizontal red solid line is the epidemic threshold, $f=0.05$ (a), $0.15$ (b), while the horizontal dotted lines are $25\%$ and $50\%$ of the mean final density of recovered without any vaccination.}\label{ws}
\end{figure}
Fig. \ref{ba} collects the results for BA[2] for the two different epidemic thresholds. As it can be expected, degree-based schemes are the most efficient in the pure BA setting. In particular, TI is the best choice in reducing the epidemic prevalence 
$\langle d_R \rangle$ by $50\%$. Our strategy (with the optimal choice $\beta =20$ and $\gamma=10$) performs very similarly at low $\langle d_V \rangle$ for both values of the epidemic threshold. However, if we want to reduce the prevalence to the $25\%$, a fast response to the outbreak is crucial, {\it i.e.} $f=0.05$. In this case, TI and LMTI are the most indicated strategies as they requires a vaccinated fraction around $10\%$. Moreover, LMTI can further reduce the epidemic prevalence for lower $\langle d_V \rangle$ than TI. On the other side, a late reaction to the epidemic ($f=0.15$) causes the difficulty in controlling the spreading, so a massive vaccination process is needed. In fact, LMTI (which is the best choice in this eventuality) requires the vaccination of at least the $25\%$ of the entire population. Instead, TI fails for $\langle d_V \rangle < 0.4$. A similar behaviour holds also in the BA[3] case, so we cease to give more details on this.
\par
\begin{figure}[!hbt]
    \centering
      \includegraphics[width=1.0\columnwidth]{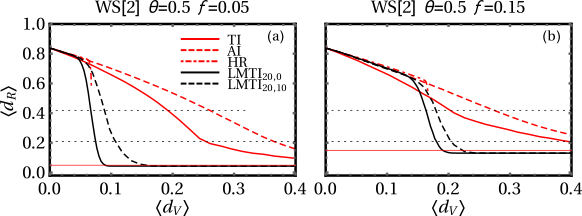}\vspace{-0.5cm}
      \caption{{The recovered mean final density $\langle d_R \rangle$ as a function of the mean fraction of vaccinated $\langle d_V \rangle$, for WS[2] with $\mathcal N=1000$ nodes and the rewiring probability $\theta=0.5$.
		The LMTI scheme is compared to TI, AI and HR immunization strategies. The horizontal red solid line is the epidemic threshold, $f=0.05$ (a), $0.15$ (b), while the horizontal dotted lines are $25\%$ and $50\%$ of the mean final density of recovered without any vaccination.}}\label{fig:ws05}
\end{figure}
\begin{figure}[!htb]
    \centering
     \includegraphics[width=1.0\columnwidth]{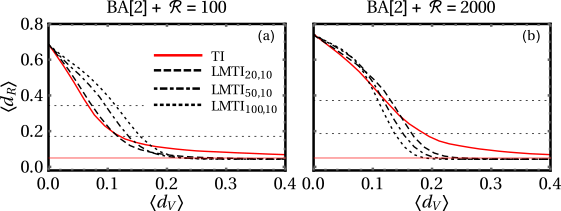}\vspace{-0.5cm}
  \caption{The recovered mean final density $\langle d_R \rangle$ as a function of the mean fraction of vaccinated $\langle d_V \rangle$, for  randomly rewired BA[2] with $\mathcal N=1000$ nodes and \mbox{$\mathcal R$=100 (a), 2000 (b)} rewiring events. 
	The LMTI scheme is compared to TI immunization strategy. The horizontal red solid line is the epidemic threshold $f=0.05$, while the horizontal dotted lines are $25\%$ and $50\%$ of the mean final density of recovered without any vaccination.}\label{rew}
\end{figure}
\begin{figure}[!htb]
     \centering
     \includegraphics[width=1.0\columnwidth]{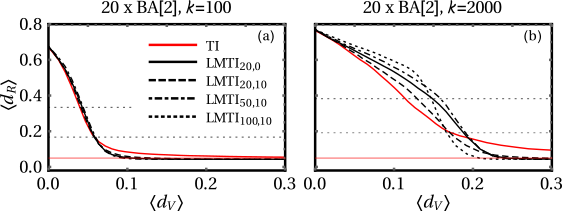}\vspace{-0.5cm}
  \caption{The recovered mean final density $\langle d_R \rangle$ as a function of the mean fraction of vaccinated $\langle d_V \rangle$,for  randomly connected BA[2] with $\mathcal N=5000$ total nodes, $m=20$ equally populated clusters  and \mbox{$ k=100$ (a), $2000$ (b)} new links.
	The LMTI scheme is compared to TI immunization strategy. The horizontal red solid line is the epidemic threshold $f=0.05$, while the horizontal dotted lines are $25\%$ and $50\%$ of the mean final density of recovered without any vaccination.}
\label{composite}
\end{figure}
In the WS setting, results are radically different, see Fig. \ref{ws} for the WS[2] and $\theta =0.1$ case. Here, TI immunization is a poor strategy when compared to LMTI and HRI. This is a consequence of the absence of nodes acting as hubs for the epidemic spreading. However, both LMTI and HRI allow to reduce the prevalence by $50\%$ for a very small number of vaccinations ($\langle d_V \rangle \lesssim 0.05$ for both values of the epidemic threshold). Most remarkably, our strategy can reduce it to $25\%$ for both values of the epidemic threshold with a vaccinated fraction lower than $10\%$ of the entire population (for comparison, AI has the same effect for $\langle d_V \rangle= 0.2 \div 0.4$). Therefore, a prompt reaction has the only effect of lowering the vaccination coverage needed to reach the aim. WS networks with different $Q$ and $\theta$ present analogous features, with the only difference that HRI dramatically worsens as the rewiring probability increases{, see Fig. \ref{fig:ws05}}. In both BA and WS cases, our strategy allows to reach the epidemic threshold and to effectively stop the epidemic.
\par
The importance of local terms in (\ref{1.2}) can be better appreciated in the BA variants. Figs. \ref{rew} and \ref{composite} collects the results for these models with the epidemic threshold $f=0.05$. In this case, we compare only TI and LMTI, the best performers in the original BA setting.\par
For partially randomized BA[2] with $\mathcal R =100$, the network keeps an approximate BA structure, so the results are very similar to the pure case. As the randomization increases, TI gradually becomes inefficient (except for small $\langle d_V \rangle$ values), so it is convenient to vaccinate nodes near the epidemic front. This is clear in the $\mathcal R=2000$ case.\par
Now, we consider randomly connected BAs with an highly clustered structure ($m=20$). If these clusters are poorly connected ($k=100$), TI and LMTI gives approximately the same performances, with the only difference that our scheme allows to stop the epidemic with a much smaller vaccinated fraction ($\langle d_V \rangle\sim 0.10$) than TI. For a much larger number of connections between the clusters ($k=2000$), the situation radically changes. In fact, the reduction of the prevalence by $50\%$ is better accomplished with TI scheme. For LMTI, the increase of local terms importance worsens the efficiency at low $\langle d_V \rangle$, but drastically improves the performance for a larger number of vaccinations.\par
This behaviour has a simple explanation. When the networks or their clusters have an approximately BA structure, nodes acting as hubs are still present. Therefore, in this case it is convenient to vaccinated nodes with higher degree. As the original structure is lost (increasing the randomization or the number of new links between the original clusters), the importance of hubs in the epidemic spreading is drastically downsized. Once that highest degree nodes are immunized, it is better to give much more importance to individuals near the epidemic front. This also explains the faster decaying of LMTI curves for increasing $\beta$ values. 
\par
\begin{figure}[!htb]
   \centering
     \includegraphics[width=1\columnwidth]{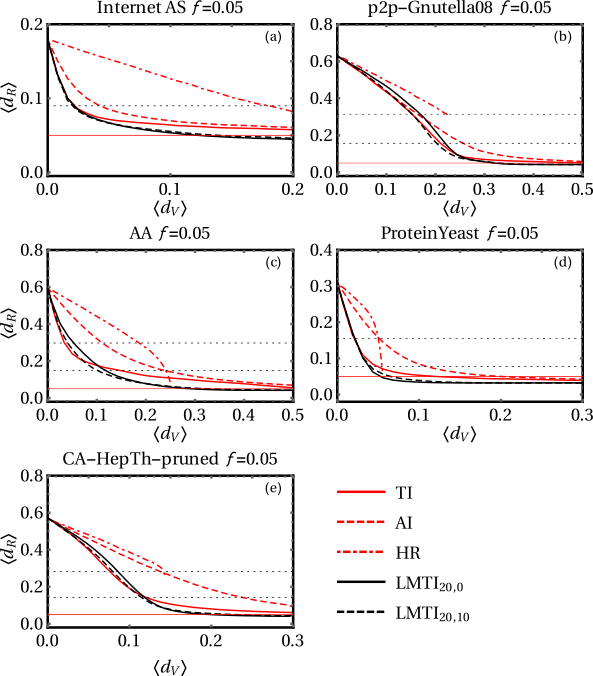}\vspace{-0.5cm}
  \caption{The recovered mean final density $\langle d_R \rangle$ as a function of the mean fraction of vaccinated $\langle d_V \rangle$ for a set of real networks (a-e). The LMTI scheme is compared to TI immunization strategy. The horizontal red solid line is the epidemic threshold $f$, while the horizontal dotted lines are $25\%$ and $50\%$ of the mean final density of recovered without any vaccination. For {\sf {Internet\_AS}} (a), the  horizontal dotted line is the $50\%$ of the mean final density of recovered without any vaccination.}\label{real}
\end{figure}

Finally, in Fig. \ref{real} we report the results for real networks. In order to halve the epidemic prevalence, we note again that TI and LMTI are the most indicated strategies and their performances are almost equivalent. In particular, TI performs slightly better only in {\sf {CA-HepTh-pruned}} and {\sf {AA}}. However, if we want to further reduce the epidemic prevalence up to $25\%$, LMTI is always the best choice. Moreover, it allows to effectively stop the epidemics for a smaller vaccinated fraction than TI. Remarkably, HRI is a rather inefficient choice also in {\sf {ProteinYeast}} and {\sf {Internet\_AS}} networks, which show a great structural resistance to the epidemics (even without any vaccination, the average size of an infection is relatively small). When compared to HRI, AI seems to be stronger, but it is comparable in efficiency to TI and LMTI only in 
{\sf {p2p-Gnutella08}}, in which it is more difficult to control the epidemic spreading (without immunization, the average size of an infection is about the $65\%$ of the entire population). This feature can be explained noting that this network is highly and uniformly connected as it presents the highest mean degree and lowest mean vertex eccentricity.
{\footnotesize
 \begin{table}[h]
 \centering
 \begin{tabular}{|c|cc|cc|}
 \hline
  & $C$ & $\ell$ & $C_{\text{R}}$ & $\ell_{\text{R}}$ \\
\hline
 \text{{\sf\small {CA-HepTh-pruned}}} & 0.28 & 5.9 & 0.0007 & 5.4 \\
\hline
 \text{{\sf\small {p2p-Gnutella08}}} & 0.020 & 4.6 & 0.0010 & 4.8 \\
\hline
 \text{{\sf\small {AA}}} & 0. & 4.4 & 0.0044 & 4.6 \\
\hline
 \text{{\sf\small {Internet\_AS}}} & 0.0096 & 3.6 & 0.00039 & 6.6 \\
\hline
 \text{{\sf\small {ProteinYeast}}} & 0.079& 6.8 & 0.0017 & 6.4 \\
 \hline
\end{tabular}
 \caption[]{
\label{tabella}
We report the global clustering coefficient $C$ and the mean distance among the nodes $\ell$ for the five real networks.
The last two columns show, as comparison, the same quantities computed for a random graph with the same number of nodes and links.}
\end{table} }

\section{Conclusions}
In this work, we have proposed a new reactive immunization strategy based on a local modification of the Targeted Immunization protocol. The aim of the local term is to actively take into account the presence of the epidemic outbreak and design  the reactive vaccination by exploiting the infection itself as a probe
of the complex network. Our proposal fits in the  framework of commonly 
very appreciated techniques using local knowledge about complex systems,  see for instance the Hebbian learning rule \cite{citeulike:500649} for an exemplary model for neural networks {and \cite{0295-5075-94-1-10002} for a detailed analysis.}
By means of explicit simulations we have compared our immunization scheme with other immunization strategies. We have shown that our protocol is a very efficient choice in all the considered cases,
allowing to stop the epidemic with a relatively small vaccinated fraction. The addition of a 
local term sensing the infection was motivated by a naive picture of the infection diffusion
valid for a regular network. Nevertheless, it is relevant also on a broad set of complex
network totally far from being regular. {We did not find a way to predict \textit{a priori} the best choice for the parameters of our score. In a purely phenomenological approach, the best free parameters are chosen empirically by looking at the performance of our scheme as $\beta$ and $\gamma$ are changed. Hopefully, a deeper investigation or the application of our score in simpler models could help to 
settle this issue.}\footnote{{Since the output of the score is an ordering of the nodes which gives the priority for the immunization, the result turns out to be robust with respect small changes of the two free parameters.}}\par
Several extension{s} of our work can be foreseen. On the theoretical  side, one can explore 
other classes of ideal networks with good theoretical control, 
like weighted or directed graphs. From the point of view of applications, 
it could be important to apply our scheme to actual specific diseases, {\em e.g.} Xylella fastidiosa, TBC and Ebola outbreaks. This will require a more realistic propagation model like the delayed SIR 
considered in \cite{agliari2013application}, and a detailed cost benefit analysis taking into account the finite resources available for
a real vaccination programme, {see for instance \cite{doi:10.1142/S0217984915501808}}. {Finally,
we remark that  our immunization scheme is clearly information-demanding, as it requires the full knowledge of the neighborhood of each node and the pattern of the epidemic at the vaccination time. 
This is rather unlikely in real situations and another natural evolution of the present work would be the study of an immunization strategy accounting the possibility of partial or corrupted information about the system. }

\bibliographystyle{JHEP}
\bibliography{ComplexNetworks-Biblio}


\end{document}